\documentclass{jfm}
\usepackage{graphicx}
\usepackage{epstopdf, epsfig}
\usepackage{xcolor}
\usepackage{comment}

\newcommand{\B}[1]{{\textcolor{red}{#1}}}

\shorttitle{Stress field in the vicinity of a bubble/sphere moving in a dilute surfactant solution}
\shortauthor{H. Kusuno, Y. Tagawa}

\title{Stress field in the vicinity of a bubble/sphere moving in a dilute surfactant solution}
\author{Hiroaki Kusuno\aff{1}\corresp{\email{kusuno\_h@kansai-u.ac.jp}}, \and Yoshiyuki Tagawa\aff{2}}

\affiliation{\aff{1}Department of Mechanical Engineering, Kansai University,
Osaka, 564-8680, Japan
\aff{2}Department of Mechanical Systems Engineering, Tokyo University of Agriculture and Technology,
Tokyo, 184-8588, Japan}

\begin{document}

\maketitle

\begin{abstract}
In this study, we experimentally investigated the stress field around a gradually contaminating bubble  as it moved straight ahead in a dilute surfactant solution with an intermediate Reynolds number ($20 < Re < 220$) and high Peclet number.
Additionally, we investigate the stress field around a falling sphere unaffected by surface contamination. 
A newly developed polarization measurement technique, highly sensitive to the stress field in the vicinity of the bubble or the sphere was employed in these experiments. 
We first validated this method by measuring the flow around a solid sphere sedimenting in a quiescent liquid at a terminal velocity. 
The measured stress field was compared with established numerical results for $Re = 120$. 
A quantitative agreement with the numerical results validated this technique for our purpose.
The results demonstrated the ability to determine the boundary layer. 
Subsequently we measured a bubble rising in a quiescent surfactant solution.
The drag force on the bubble, calculated from its rise velocity, was set to transiently vary from that of a clean bubble to a solid sphere within the measurement area.
With the intermediate drag force between clean bubble and solid sphere, the stress field in the vicinity of the bubble front was observed to be similar to that of a clean bubble, and the structure near the rear was similar to that of a solid sphere.
Between the front and rear of the bubble, the phase retardation exhibited a discontinuity around the cap angle at which the boundary conditions transitioned from no-slip to slip, indicating an abrupt change in the flow structure.
A reconstruction of the axisymmetric stress field from the phase retardation and azimuth obtained from polarization measurements experimentally revealed that stress spikes occur around the cap angle.
The cap angle (stress jump position) shifted as the drag on the bubble increased owing to surfactant accumulation on its surface. 
Remarkably, the measured cap angle as a function of the normalized drag coefficient quantitatively agreed with the numerical results at intermediate $Re = 100$ of \citet{Cuenot1997}, exhibiting only a slight deviation from the curve predicted by the stagnant cap model at low $Re$ (creeping flow) proposed by \citet{Sadhal1983a}.

\end{abstract}


\section{Introduction}\label{sec:introduction}

The rising velocity of a bubble in a surfactant solution can be reduced to half its velocity in pure water \citep[]{Magnaudet2000,Takagi2011}, which is the same as that of a solid sphere.
This behavior can be traced back to mechanisms first elucidated by \citet{Frumkin1947} and \citet{Levich1962}. 
As bubbles rise at a high Peclet number $Pe$ ($=dU/D$, where $d$ is the bubble diameter, $U$ is the bubble velocity, and $D$ is the diffusion coefficient), surfactants at the bubble surface move from the front to the rear owing to surface advection. 
Consequently, a higher surfactant concentration occurs at the rear of the bubble, creating a gradient of surface tension along the bubble surface. 
Unlike the typical gas-liquid boundary conditions that have zero shear stress, here, the stress occurs owing to the surface tension gradient, known as the Marangoni effect. 
This modifies the flow around the bubble, increasing its drag. 
In many scenarios, even a minimal surfactant contamination of the order of a few p.p.m. can cause the  drag force on the bubble to be comparable to that of a rigid sphere, as reported by \citet{Harper1972} and \citet{Clift1978}.

The variables influencing the velocity of a bubble were studied by \citet{Bachhuber1974}, who observed that bubbles for $Re \:(=\rho dU/\mu)<$ 40 in water begin their ascent with the velocity of a sphere under a free-slip boundary condition, but as the bubble moved upwards, it rises with the velocity of a solid sphere. 
This indicates that the surfactant concentration on the bubble surface increases with distance traveled, resulting in an increase in drag. 
Notably, \citet{Zhang2001} showed that at low surfactant concentrations, the distance to reach the terminal velocity of a solid sphere can span several meters because of the slow pace of adsorption and desorption of the surfactant.

The aforementioned mechanism is modeled in the so-called stagnant cap model \citep{Savic1953,Davis1966,Harper1982,Sadhal1983a}. 
In this model, two boundary conditions are considered, separated by the cap angle $\theta_C$ ($\theta_C$ corresponds to the leading edge of the stagnant region and is defined here from the front stagnant point): zero shear stress in the bubble front up to $\theta_C$ and a non-slip condition afterwards.
\citet{Sadhal1983a} formulated a drag equation as a function of cap angle in creeping flow, demonstrating significant drag force variations within a specific angle range ($60^\circ < \theta_{C} < 150^\circ$). 
The flow structure in the vicinity of a bubble has been investigated primarily using numerical analysis \citep[e.g.,][]{Cuenot1997,Takemura1999,Ponoth2000,Dani2006,Dani2022,pesci2018computational,Kentheswaran2022,Kentheswaran2023}. 
Intriguingly, many studies reported that this drag increase as a function of the cap angle remains consistent even at intermediate Reynolds numbers ($Re$), except 
 for the specific angle range where the slope is slightly steeper than that predicted by the model \citep{Cuenot1997}.  

The change in the boundary conditions owing to the surfactant has other important effects in addition to its effect on the velocity of the bubble.
\citet{Tagawa2014} experimentally investigated the effects of surfactants on the path instability of a single bubble.
They observed that when $Re$ is fixed at 400, the bubble motion transitions from rectilinear to helical spiral and further to zigzag motion with an increase in the drag coefficient, corresponding to increased surfactant concentration.
This mechanism is explained by changes in vorticity generation owing to altered boundary conditions and vorticity stretching/tilting.
Therefore, the extent and location of boundary condition changes, and the effect of these changes on the flow field in the vicinity of an object, should be clarified.


However, experimental observations on stress fields near a bubble/sphere moving in a dilute surfactant solution are quite limited. 
The measurement of a flow field in the vicinity of a bubble remains a significant challenge owing to the thin boundary layer on the bubble surface.
The classical visualization technique, e.g., particle image velocimetry (PIV) is particularly difficult because it uses micron-sized fluid tracers.
More importantly, to locate the location of the cap angle, we would prefer to visualize the stress field.
Although fluid stresses can be estimated from PIV data, the obtained stress data are often noisy because it requires the calculation of spatial derivatives of the velocity field.

In this study, we experimentally examined the stress field around a rising bubble in a dilute surfactant solution using photoelastic measurement, a recently developed flow visualization method employing nano-sized rods \citep{nakamine2024flow}.
To validate this technique, we first analyzed the stress field around a solid sphere sinking in a quiescent liquid with terminal velocity, comparing our results with well-established numerical simulation.
This comparison also served to characterize the stress field simulating a fully contaminated bubble. 
Subsequently, we investigated a bubble rising rectilinearly in a quiescent surfactant solution at intermediate $Re$ ($20 < Re < 220$) and high Peclet numbers $Pe = O(10^3-10^5)$.
By observing bubbles at various distances traveled, we characterized the stress field near the bubble surface, where the boundary condition transitioned from a clean to a contaminated state.
The cap angle was then estimated from the experimental result.
Finally, we discuss its relevance described in the stagnant cap model previously reported and previous numerical study by \citet{Cuenot1997}.

\section{Experimental and numerical methods}\label{sec:experimental_method}
\subsection{Experimental methods}\label{subsec:experimentalsetup}
\subsubsection{Experimental setup}\label{subsubsec:experimentalsetup}

\begin{figure}
    \centerline{\includegraphics[width=1\linewidth]{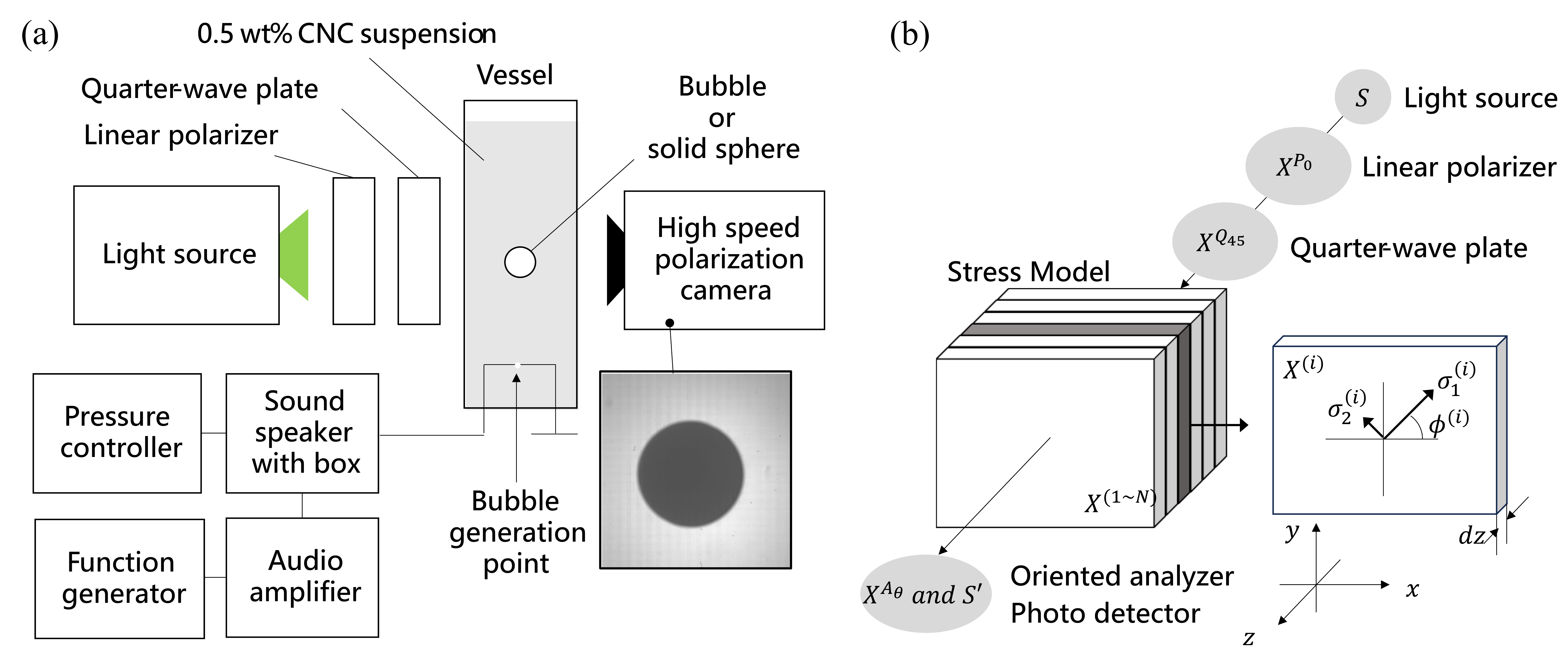}}
    \caption{The schematic of the experimental setup. (a) Overview. (b) The schematic diagram of measurement principle.}
\label{fig:ex_set}
\end{figure}

We observed the flow structure around a bubble and a solid particle in a cellulose nanocrystal (CNC) suspension that exhibits photoelasticity \citep{Lane2022,nakamine2024flow,worby2024examination}.
Flow birefringence occurs when non-spherical nanoparticles align in a certain orientation under the effects of shear \citep{calabrese2021b}.
By controlling the polarization state before flow birefringence and measuring the polarization state after, we can determine the phase retardation $\Delta$ and azimuth $\phi$.
According to the stress-optic law, $\Delta$ and $\phi$ correspond to the magnitude and component of stress, enabling the calculation of the integrated stress within the flow field.

Figure \ref{fig:ex_set} shows a schematic of the experimental setup.
Back light generated by a light source (SOLIS-525C, Thorlabs) of a wavelength $\lambda$ of 520 nm was circularly polarized through a linear polarizer and a quarter-wave plate.
The circularly polarized light passed through the flow around the object, resulting in the emission of elliptically polarized light with $\Delta$ and $\phi$.
A high-speed polarization camera (CRYSTA PI-1P, Photron Ltd.) was used to measure $\Delta$ and $\phi$.
The camera had linear polarizers whose angles differed by $45^\circ$ at neighboring pixels \citep{onuma2014development}.
Using the intensity distribution of the four pixels set (superpixels), we derived the phase retardation and azimuth.
Applying the four-step phase-shifting method, we described $\Delta$ and $\phi$ using the intensities measured through each linear polarizer, denoted as $I_0, I_{45}, I_{90}$, and $I_{135}$ \citep{Lane2022,Yokoyama2023,nakamine2024flow}:
\begin{equation}
\Delta = \frac{\lambda}{2\pi}\sin^{-1}{\frac{\sqrt{(I_{90}-I_0)^2 + (I_{45}-I_{135})^2}}{I/2}}\;,
\label{eq:intensities to delta}
\end{equation}
\begin{equation}
\phi = \frac{1}{2}\tan^{-1}{\frac{I_{90}-I_0}{I_{45}-I_{135}}}\;,
\label{eq:intensities to phi}
\end{equation}
where $I$ is the sum of the four intensities, i.e., $I_0, I_{45}, I_{90}$, and $I_{135}$.
The image resolution for the phase retardation and azimuth images was 512 $\times$ 512 pixels.
The camera captured images at 2,000 fps and 10–20 $\rm{\mu m/pix}$.
A time-averaged values over 10 frames was applied to all experimental data.


We used a CNC (Cellulose Lab Ltd.) suspension.
A 0.5 wt$\%$ of CNC and 0.4 ppm of Triton X-100 were mixed with ultrapure water using a magnetic stirrer and then sonicated for 15 min using a homogenizer (UX-300, Mitsui Electric Co., Ltd).
This treatment increases the transparency of a suspension, and the suspension behave as a Newtonian fluid.
This allows for comparisons with numerical analysis of Newtonian fluids. 
The viscosity of the CNC suspension $\mu$ was $1.7\rm mPa\cdot s$ at $10^\circ\rm C$ measured using a rheometer (MCR302, Anton Paar Co. Ltd.).

The CNC suspension was filled into a vessel (90 mm $\times$ 90 mm $\times$ 450 mm, glass), in which the effects of the wall on the flow stracture around an object rising/falling at the center of the vessel could be neglected \citep{Clift1978}.
A single bubble was generated from the bottom of the vessel using a sound speaker (FF125K, Fostex) \citep{Kusuno2019}.
The sound speaker was activated using an input signal via a function generator (WF1974, NF Corp.) and an audio amplifier (AP20d, Fostex).
The static pressure at the outlet of the bubble generator was controlled using a pressure controller (PACE 5000, Baker Hughes).
The measurement height varied between 5 and 400 mm from the outlet of the bubble generator.
We observed a bubble rising in a rectilinear path within the measurement range, indicating that the flow around the bubble was axisymmetric.
The $Re$ was within the range of $20 < Re < 220$.

To verify the performance of the polarization measurement, we used a solid sphere ($1770\rm \:kg/m^3, 6 \:mm \:in\:diameter, Plastic$).
The sphere was quietly dropped from the surface of the suspension.
We confirmed that no gas was adsorbed at the solid-liquid interface.
The camera was positioned in an area with a sphere's falling distance of 120 mm, which was sufficiently long to assume a steady-state condition to be reached at $Re = 120$.

\subsubsection{Integrated photoelasticity}\label{sec:integrated photoelasticity}
Considering a two-dimensional stress field of infinitesimal thickness $dz$ around an object, the phase retardation $\Delta$, azimuth $\phi$, and stress $\sigma$ are related by
\begin{equation}
\Delta\cos 2\phi = C(\sigma_{xx}-\sigma_{yy})dz\;,
\label{eq:Retadation1}
\end{equation}
\begin{equation}
\Delta\sin 2\phi = 2C\sigma_{xy}dz\;,
\label{eq:Retadation2}
\end{equation}
where $\sigma_{xx}$, $\sigma_{yy}$, and $\sigma_{xy}$ are the stress components in the Cartesian coordinate system shown in figure \ref{fig:ex_set}(b), and $C$ is the material-specific constant.
Because we consider a two-dimensional stress field, the stress component in the optical axis direction, i.e., $z$-direction, does not affect the photoelastic measurement results.
However, our experimental results for the phase retardation and azimuth were three-dimensional stress fields in which the fluid stress field was integrated along the optical path.
Here, we adopted the method of integrated photoelasticity \citep{ramesh2000digital}, where the stress field is sliced into a collection of virtual thin plats, which are sufficiently thin to be assumed as a two dimensional stress fields (figure \ref{fig:ex_set}(b)).
In this case, the polarization state through the 3D stress field can be calculated by multiplying the Stokes parameter $\mathbf{S}$ by the Mueller matrix $\mathbf{X^{(i)}}$ corresponding to each plate
\begin{equation}
\mathbf{S'}=\mathbf{X^{(A_\theta)}X^{(N)}...X^{(i)}....X^{(2)}X^{(1)}X^{(Q_{45})}X^{(P_0)}S}\;,
\label{eq:Stokes}
\end{equation}
where $\mathbf{X^{(i)}}$ is the number of the plate from which the fluid stress field is sliced (including the linear polarizer $\mathbf{X^{(P_0)}}$, quarter-wave plate $\mathbf{X^{(Q_{45})}}$, and oriented analyzer $\mathbf{X^{(A_\theta)}}$) and is expressed in terms of phase retardation and azimuth \citep{nakamine2024flow}.
\begin{equation}
\setlength{\arraycolsep}{0pt}
\renewcommand{\arraystretch}{1.3}
\mathbf{X^{(i)}}= \left[
\begin{array}{ccccc}
  1 & 0 & 0 & 0 \\
0 & 1-(1-\cos\Delta^{(i)})\sin^2 2\phi^{(i)} & (1-\cos\Delta^{(i)})\sin 2\phi^{(i)}\cos 2\phi^{(i)} & -\sin\Delta^{(i)}\sin 2\phi^{(i)} \\
0 & (1-\cos\Delta^{(i)})\sin 2\phi^{(i)}\cos 2\phi^{(i)} & 1-(1-\cos\Delta^{(i)})\cos^2 2\phi^{(i)} & \sin\Delta^{(i)}\cos 2\phi^{(i)} \\
0 & \sin\Delta^{(i)}\sin 2\phi^{(i)} & -\sin\Delta^{(i)}\cos 2\phi^{(i)} & \cos\Delta^{(i)}  \\
\end{array}  \right] .
\label{eq:Mueller}
\end{equation}
The Mueller matrices $\mathbf{X^{(i)}}...\: \mathbf{X^{(N)}}$ are calculated using Eqs.~(\ref{eq:Retadation1}),~(\ref{eq:Retadation2}), and~(\ref{eq:Mueller}), based on stress field obtained from the numerical results explained in $\S$ \ref{sec:numerical methods}.
Because the numerical results are velocity-pressure fields, viscous stresses are obtained from the velocity fields and then transformed into integrated phase retardation and azimuth using the Stokes parameters and Mueller matrices.
Finally, using Eq. (\ref{eq:Stokes}), we can obtain the integrated phase retardation and azimuth, which can be directly compared with experimental results \citep[for more details, see][]{Yokoyama2023}.

\subsubsection{Surfactant effects}\label{sec:surfactants}
In aqueous surfactant solutions, the drag coefficient of the bubble changes from that with free-slip condition to that of non-slip condition \citep{Cuenot1997}.
At high Peclet numbers and concentrations below the critical micelle concentration (CMC), surfactants on the surface are transported towards the rear of the rising bubble, resulting in a concentration gradient along the surface.
Consequently, the surface tension gradient induces the shear stress.

In this study, Triton X-100 was used, and we set the concentration to 0.4 ppm (7$\times10^{-4}$ mol$/\rm m^3$).
Because the concentration in this study was much smaller than the CMC of Triton X-100 (0.23 mol/$\rm m^3$), micelles did not form.
Furthermore, when the Peclet number is of the order of $10^3$ or higher, advection occurs on a much shorter timescale than diffusion.
Under such conditions, the sudden change in surfactant concentration on the surface, or stress jump, is determined by the advection of surfactant and the kinetic balance of adsorption and desorption.

Adsorption/desorption models of interfaces have been proposed previously \citep{Levich1962, Lin1990}.
Following Langmuir adsorption kinetics, the surfactant exchange $j_s$ can be expressed by the following equation:
\begin{equation}
j_s=k_aC_s(\Gamma_\infty-\Gamma)-k_d\Gamma\;.
\label{eq:Langmuir}
\end{equation}
where $C_s$ is the sublayer concentration, $\Gamma_\infty$ is the maximum adsorbed surface concentration, $\Gamma$ is the surface concentration, $k_a$ is the adsorption constant, and $k_d$ is the desorption constant.
For Triton X-100, $\Gamma_\infty$ = 2.9$\times\rm10^{-6}$ mol/$\rm m^2$, $k_a$ = 50 $\rm m^3/mol\cdot s$, and $k_d$ = 3.3$\rm\times10^{-2}s^{-1}$ \citep{Lin1990}.
The equilibrium interfacial concentration $\Gamma_{max}$ at $j_s = 0$ is
\begin{equation}
\Gamma_{max} = \frac{k_aC_s}{k_aC_s+k_d}\Gamma_\infty\;.
\label{eq:eq_concentration}
\end{equation}
For initially clean interfaces $(\Gamma = 0)$, the equilibrium surface concentration can be scaled using the following equation:
\begin{equation}
\Gamma_{max} = \int_0^{t_{eq}}j_s(t)dt\sim k_aC_s\Gamma_\infty t_{eq}\;,
\label{eq:eq_concentration2}
\end{equation}
where $t_{eq}$ is the time required to reach the equilibrium concentration.
From Eqs. (\ref{eq:eq_concentration}) and (\ref{eq:eq_concentration2}), the order of $t_{eq}$ is $(k_aC_s+k_d)^{-1}$.
For a bubble rising velocity of 260 $\rm mm/s$ with a radius of 0.73 $\rm mm$, the distance required to reach equilibrium is $z_{eq} \sim t_{eq}U \sim 1\rm\:m$. 
Therefore, we can infer that the velocity decreases gradually in the measurement region used in this study.
Note that in this study, the time-averaged value over 10 frames $(5\rm\:ms)$ was used.
Because the distance travelled by the bubble during this period was $O(10^{-4})\rm\:m$, which was sufficiently short compared with the characteristic distance $O(1)\rm\:m$ over which interfacial adsorption varies, the effect of averaging was not considered to be significant.

\subsection{Numerical methods}\label{sec:numerical methods}
A numerical simulation of the flow around the solid sphere was performed using the open source code Basilisk \citep{Popinet2015} to validate the phase retardation and azimuth obtained from the experimental results, which corresponded with the fluid stress field.
A flow around a fixed sphere with uniform inflow was computed.
The equations of motion and divergence-free conditions are expressed as
\begin{equation}
\frac{\partial \mathbf{u}}{\partial t} + \bnabla\cdot(\mathbf{uu})
=\frac{1}{\rho}
\left(-\bnabla p +\bnabla\cdot 2\mu\mathbf{D}
\right)\;,
\label{eq:NS}
\end{equation}
\begin{equation}
\bnabla\cdot\mathbf{u} = 0\;,\\
\label{eq:Cont}
\end{equation}
where $\mathbf{D}=(\bnabla\mathbf{u}+(\bnabla\mathbf{u})\rm ^T)/2 $ is the strain-rate tensor.
The sphere was described using an embedded boundary method.
For the solid sphere, the boundary condition was non-slip. 
The grid size near the interface was 1/128 of the diameter, which was sufficient to resolve the boundary layer of the order of $O(Re^{-0.5})$ in thickness \citep[e.g.][]{lamb1945}.
The computational domain was 40 times larger than the diameter.
The effect of boundary conditions at the wall of the computational domain could be neglected.
The length of the standing eddy at $Re$ = 100 was determined to be 0.82$d$ with a separation angle of 128$^\circ$ and that at $Re$ = 300 was determined to be 1.33$d$, with a separation angle of 112$^\circ$, which agreed with previous studies \citep{pruppacher1970some,Clift1978,magnaudet1995}, thus validating the numerical results.

\section{Results and discussion}\label{sec:result}
\subsection{Flow around a solid sphere}\label{sec:solid}


\subsubsection{Measurement validation: stress field around a solid sphere at $Re=120$}\label{sec:verification}

\begin{figure}
\centerline{\includegraphics[width=1\linewidth]{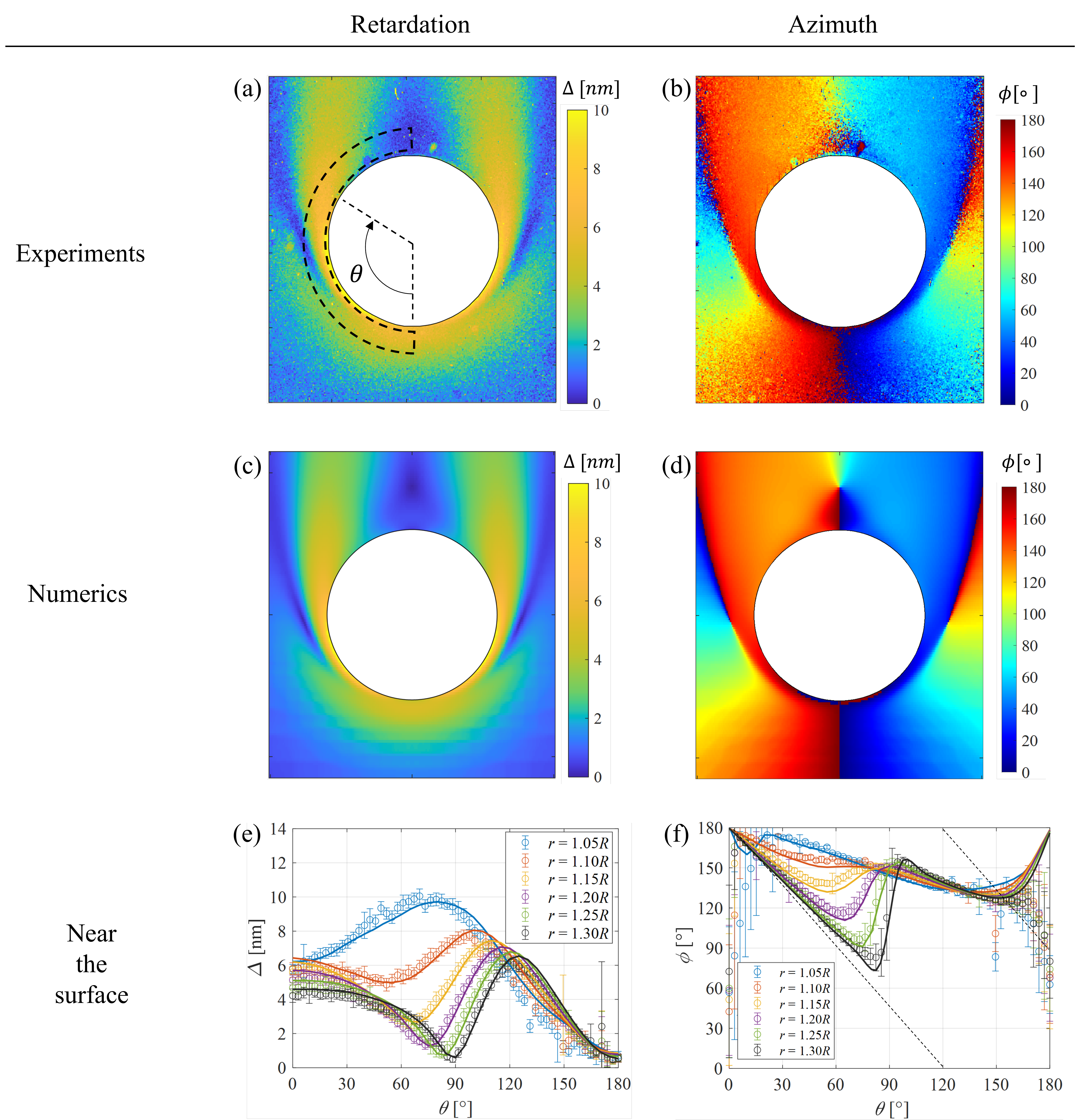}}
    \caption{Flow surrounding a solid sphere at $Re$ = 120. (a) Experimental phase retardation. (b) Experimental azimuth. (c) Numerical phase retardation. (d) Numerical azimuth. (e) Phase retardation near the surface. (f) Azimuth near the surface. The dashed line of (a) is the region of $r = 1.05R$ to $1.30R$.}
\label{fig:solid}
\end{figure}

In this subsection, we validate the photoelastic measurement technique by comparing experimental results with established numerical simulations. 
This comparison was conducted both qualitatively and quantitatively to demonstrate the method’s accuracy and reliability in capturing stress fields around bluff bodies moving through a quiescent liquid. 
Specifically, we focused on the flow around a solid sphere at $Re = 120$, representing an intermediate $Re$ regime similar to that of bubbles.\\ \indent
Figures \ref{fig:solid}(a) and (b) show the experimental results for phase retardation and azimuth fields around a solid sphere falling in a CNC suspension at $Re$=120. 
The white region corresponds to the sphere itself. 
The phase retardation, which is proportional to the principal stress difference, and the azimuth, which reflects the stress component alignment, were observed during the motion of the sphere. \\ \indent
Figures \ref{fig:solid}(c) and (d) present the corresponding numerical simulations for the same system. 
A direct comparison revealed reasonable agreement between the experimental and numerical results, indicating the accuracy and reliability of the polarization technique. This validation highlighted the potential of this method in capturing stress fields quantitatively.\\ \indent
Figures \ref{fig:solid}(e) and (f) depict detailed phase retardation and azimuth profiles in the near-surface region, specifically from $r = 1.05R$ to $1.30R$. 
Remarkably, the experimental measurements exhibited excellent agreement with numerical results within this range for $20^\circ < \theta < 160^\circ$, confirming the quantitative accuracy of the stress-optic law as applied through polarization techniques. 
This agreement also demonstrated the capability of this method to accurately resolve stress fields in the boundary layer, whose thickness is approximately $\delta/R \sim \sqrt{1/Re} \sim O(10^{-1})$.
The ability to resolve stress fields in this region supports analysis of flow behavior around the sphere, including detailed discussions on boundary layer dynamics and wake structures as discussed later in the following sections.\\ \indent
In areas close to the stagnation points (both front and rear, i.e. $\theta \sim 0^\circ$ and $180^\circ$), experimental azimuth measurements exhibited larger errors. 
This discrepancy resulted primarily from two factors: (1) the azimuth angle shifted sharply from $180^\circ$ to $0^\circ$ across the symmetry axis, where shear stress changed sign, and (2) the phase retardation in these regions was minimal, resulting in reduced measurement accuracy. In addition, data acquisition techniques, particularly at the interface ($r < 1.05R$), were critical for further reducing errors by noise and data resolution near the interface owing to pixel shadowing and fluctuations.\\  \indent
Despite the inherent challenges, the polarization technique demonstrated its capability to capture key features of the stress field. 
These results established its utility for analyzing particle-laden flows, including the flow around spheres and bubbles examined in this study.
Furthermore, the findings suggest the potential to extend the method to more complex multiphase and non-Newtonian flow systems, subject to further investigation and refinement.

\subsubsection{Physical insights into the stress field around a sphere}
\label{sec:flowbased}

\begin{figure}
\centerline{\includegraphics[width=0.7\linewidth]{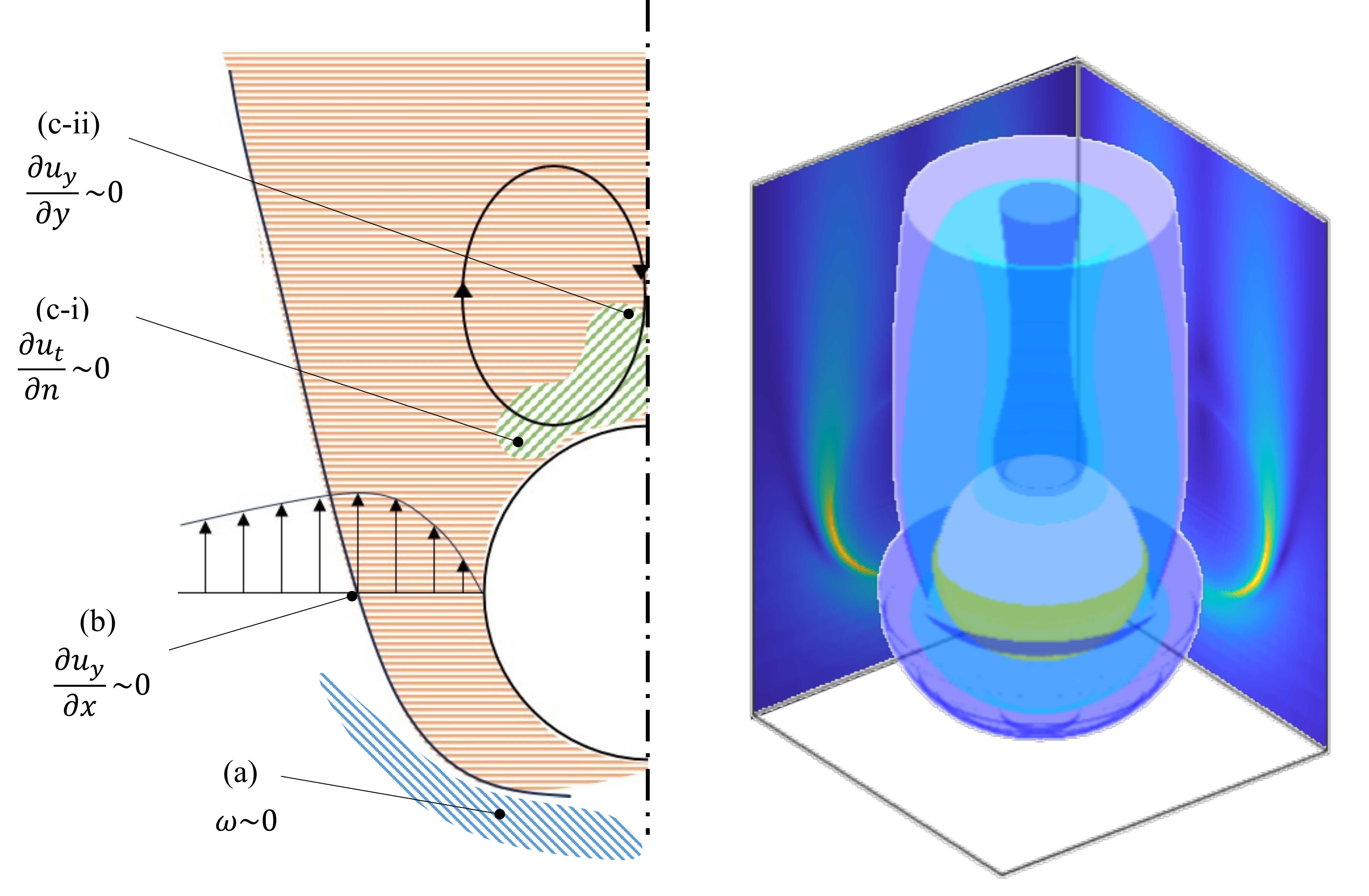}}
   \caption{(Left) Schematic of the flow field around a solid sphere at $Re = 120$. Key features of the flow are labeled: (a) mushroom-shaped region of vanishing vorticity ($\omega \sim 0$), (b) region where the normal stress gradient is negligible ($\partial u_y / \partial x \sim 0$), and (c) regions in the wake where (c-i) the tangential stress gradient at the surface vanishes ($\partial u_t / \partial n \sim 0$) and (c-ii) normal stress gradient is nearly zero ($\partial u_y / \partial y \sim 0$). (Right) 3D visualization of the flow structure around the sphere, showing the boundary layer, wake, and standing eddy formation. Experimental results related to this flow field are shown in figure~\ref{fig:solid}.}
\label{fig:pontie}
\end{figure}

\begin{figure}
\centerline{\includegraphics[width=1.05\linewidth]{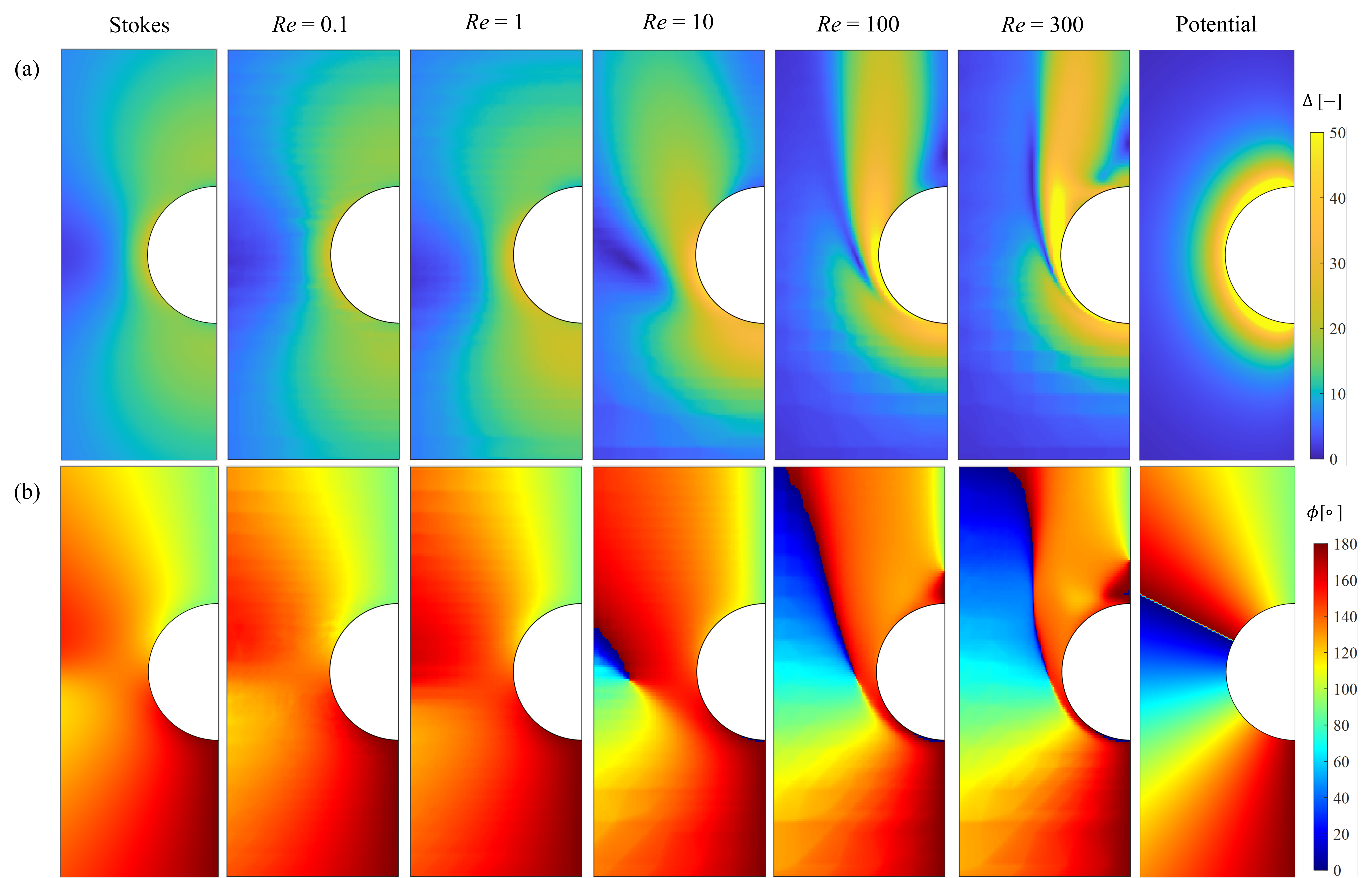}}
    \caption{Numerical results of flow around a solid sphere for various Reynolds numbers ($Re = 0.1, 1, 10, 100, 300$) and analytical results for two limiting cases (Stokes flow and potential flow). (a) Phase retardation $\Delta$ and 
    (b) Azimuth $\phi$ distribution. These results illustrate the evolution of stress fields and their dependence on $Re$ and flow regimes.}
\label{fig:all_numerical}
\end{figure}

In this section, we explore the physical characteristics of the stress field around a solid sphere, based on the experimental validation from the previous section (\S\ref{sec:verification}). 
Figure~\ref{fig:pontie} provides a schematic and 3D illustration of the flow structure around the sphere at $Re = 120$, highlighting key features such as the boundary layer, wake, and standing eddy formation. 
To deepen our understanding, figure~\ref{fig:all_numerical} systematically presents stress fields at various $Re$, including two limiting cases: Stokes and potential flows. 
This comparison enables us to explore the evolution of the stress field across different regimes, from low to intermediate $Re$ and potential flow limits.

To enhance readability and understanding, we first examine the stress fields across different $Re$ using figure~\ref{fig:all_numerical}. 
For the potential solution, the phase retardation decreases with increasing distance from the sphere's surface and exhibits a slight dependence on the angular coordinate, $\theta$. 
In contrast, the azimuth profile strongly depends on $\theta$ rather than on the radial distance, $r$, aligning with the potential flow velocity distributions, $u_r = U(1 - R^3 / r^3)\cos \theta$ and $u_\theta = -U(1 + R^3 / 2r^3)\sin \theta$.
In Stokes flow, where the velocity in the fluid are $u_r = U(1 + R^3/2r^3- 3R / 2r)\cos \theta$ and $u_\theta = -U(1 - R^3 / 4r^3 + 3R/4r)\sin \theta$, the phase retardation also decreases with increasing distance, but the azimuth is influenced by both $\theta$ and $r$, resulting in a profile that clearly deviates from the potential solution. 
The phase and azimuth profiles in Stokes flow align well with behavior observed in the low $Re$ regime ($Re \leq 1$). 
At higher $Re$ ($10 \leq Re$), a boundary layer develops around the sphere, and the azimuth profile in the outer flow begins to resemble that of the potential solution.
At intermediate $Re$ (10, 100, 300), the flow structure transitions from a viscous-dominated regime to an inertial-dominated regime, with noticeable deviations from symmetry in the phase retardation fields. These trends provide a framework for understanding the specific characteristics of the flow at $Re = 120$, which are discussed in detail in the following paragraphs.

The flow field at $Re = 120$ is depicted in figure~\ref{fig:pontie}, highlighting key regions of interest: (a) the mushroom-shaped phase retardation at the front of the sphere, observed despite not being the area of maximum stress, with negligible vorticity ($\omega \sim 0$); (b) the boundary layer, extending until the normal stress gradient becomes minimal ($\partial u_y / \partial x \sim 0$); and (c-i) and (c-ii), corresponding to regions in the wake in which tangential and normal stress gradients vanish. In the following, we discuss the distinct characteristics of each region:\\ \indent
In the region shown in figure~\ref{fig:pontie}(a), which corresponds to the area in front of the sphere and outside the boundary layer, the flow exhibits characteristics of potential flow, with negligible vorticity ($\omega \sim 0$). 
The azimuth profile shown in figure~\ref{fig:solid}(b)(d) aligns with the theoretical potential flow solution, and the phase retardation decreases smoothly with distance from the sphere. 
However, this alone does not explain the mushroom-shaped phase retardation pattern observed at $Re = 120$, which requires considering the combined effects of potential flow and the boundary layer.\\ \indent
In the outer region (figure~\ref{fig:pontie}(a)), the flow follows potential flow behavior, characterized by negligible vorticity and a stress distribution primarily dependent on the angular coordinate, $\theta$, rather than the radial distance, $r$. 
This corresponds with the velocity distribution predicted for potential flow, where phase retardation decreases gradually with increasing distance from the sphere, setting a baseline for the retardation profile in the outer flow region.\\ \indent
Closer to the sphere, within the boundary layer, the no-slip condition at the solid surface causes the velocity to sharply drop to zero, creating steep velocity gradients and significant vorticity. 
This leads to a pronounced increase in phase retardation near the sphere’s surface. 
At the boundary layer’s edge, where the transition to potential flow occurs, the velocity gradient reverses, reducing viscous stress and causing the phase retardation to drop abruptly. 
This sharp transition at the boundary layer edge significantly contributes to the mushroom-shaped retardation pattern.\\ \indent
Thus, the combined effects of smooth potential flow behavior in the outer region and the steep velocity gradients within the boundary layer result in the formation of the mushroom-shaped phase retardation at $Re = 120$. 
The distinct drop in retardation at the boundary layer edge reveals the interplay between these two flow regimes, highlighting the unique structure of the stress field around the sphere at intermediate $Re$.

In the region shown in figure~\ref{fig:pontie}(b), the boundary layer forms as a result of the no-slip condition at the sphere's surface. 
Within this region, phase retardation increases significantly owing to steep velocity gradients near the surface, whereas the azimuth profile reflects the orientation of the stress components induced by these gradients. 
A large phase retardation, indicating high stress, is observed from the front stagnation point to the equator of the sphere. 
In this region, the azimuth remains relatively constant, with values ranging from $135^\circ$ to $180^\circ$ on the left side of the sphere and $0^\circ$ to $45^\circ$ on the right side. 
This distinction in the azimuth profile highlights the difference in flow structure between the near-solid region and the surrounding flow, marking the boundary layer as a region dominated by vorticity (or viscous stress).\\ \indent
At the boundary layer's edge, phase retardation decreases sharply, eventually becoming nearly zero as the transition to potential flow occurs. 
This transition is accompanied by a reversal in the velocity gradient, which reduces the viscous stress.
The azimuth profile further illustrates this transition, with azimuth values of approximately $135^\circ$ near the solid surface and $45^\circ$ immediately outside the boundary layer. 
This trend is particularly pronounced at higher $Re$, where the thinner boundary layer sharpens the velocity gradient and stress distribution, emphasizing the difference between the near-solid and outer flow regions.\\ \indent
In figure~\ref{fig:pontie}(b)--(c), areas of significant phase retardation do not extend to the rear stagnation point but instead spread downstream, forming the wake.
Similarly, regions with characteristic azimuth extend into the wake. 
Near the rear stagnation point, phase retardation is minimal, and the azimuth is $180^\circ$ (or $0^\circ$). 
However, as the distance from the rear stagnation point increases (approximately $0.25d$ from the solid surface), the azimuth transitions to $90^\circ$. 
The region where the azimuth remains $180^\circ$ represents part of the standing eddy but does not encompass its entire structure.\\ \indent
Overall, the unique interplay between the steep velocity gradients in the boundary layer and the transition to potential flow at the boundary layer edge contributes to the observed stress field characteristics. The phase retardation and azimuth profiles, particularly their transitions, reflect the underlying flow physics, emphasizing the role of vorticity and velocity gradient reversals in defining the boundary layer and wake structure.\\ \indent
In the region shown in figure~\ref{fig:pontie}(c-i) and (c-ii), the wake exihibits minimal phase retardation, with the azimuth transitioning from $180^\circ$ near the rear stagnation point to $90^\circ$ further downstream.
These regions highlight the formation of a standing eddy, whose size and intensity are determined by the $Re$, as depicted in figure~\ref{fig:all_numerical}. 
The stress gradients in this wake region reflect the transition from the recirculating flow behind the sphere to the external flow. 
Note that determining the presence and size of the standing eddy purely from phase retardation measurements is challenging because the retardation values in this region are small, as shown in figures~\ref{fig:solid} and \ref{fig:all_numerical}. Instead, the azimuth profile provides more reliable insights into the structure of the standing eddy.\\ \indent
First, consider the separation point on the solid surface (figure~\ref{fig:pontie}(c-i)), where tangential stress becomes approximately zero as the velocity gradient reverses before and after separation. 
In the presence of a standing eddy, the azimuth increases with $\theta$. 
Next, in the standing eddy region along the symmetry axis (figure~\ref{fig:pontie}(c-ii)), the azimuth transitions from $180^\circ$ near the rear stagnation point to $90^\circ$ further downstream. 
The distance from the solid surface to this azimuth transition point is approximately $0.24d$, which does not represent the full length of the standing eddy (about $0.94d$ at $Re = 120$). 
This transition corresponds to the condition $\partial u_y / \partial y = 0$ in the $(x, y)$ coordinate system, indicating the location of maximum velocity within the standing eddy. 
If no standing eddy is present, then $\partial u_y / \partial y < 0$ behind the object, suggesting that the azimuth profile can be used to infer the presence or absence of a standing eddy.\\ \indent
These results suggest that the narrow region between the separation point (c-i) and the maximum velocity point in the standing eddy (c-ii) is critical for understanding the stress profile in the wake. This region provides valuable insights into the interplay between recirculating flow and the transition to the external flow.\\ \indent
As demonstrated above, analyzing the phase retardation and azimuth provides valuable insights into distinguishing critical flow regions, such as the boundary layer and wake, and understanding their underlying structures. 
These measurements enable us to link stress distributions with flow dynamics, revealing the intricate interplay between viscous and inertial effects. 
While pressure visualization remains a challenge for future studies, the square of the phase retardation correlates with viscous energy dissipation, offering a potential pathway to relate stress field measurements to pressure drop and overall flow behavior.

\begin{figure}
    \centerline{\includegraphics[width=.7\linewidth]{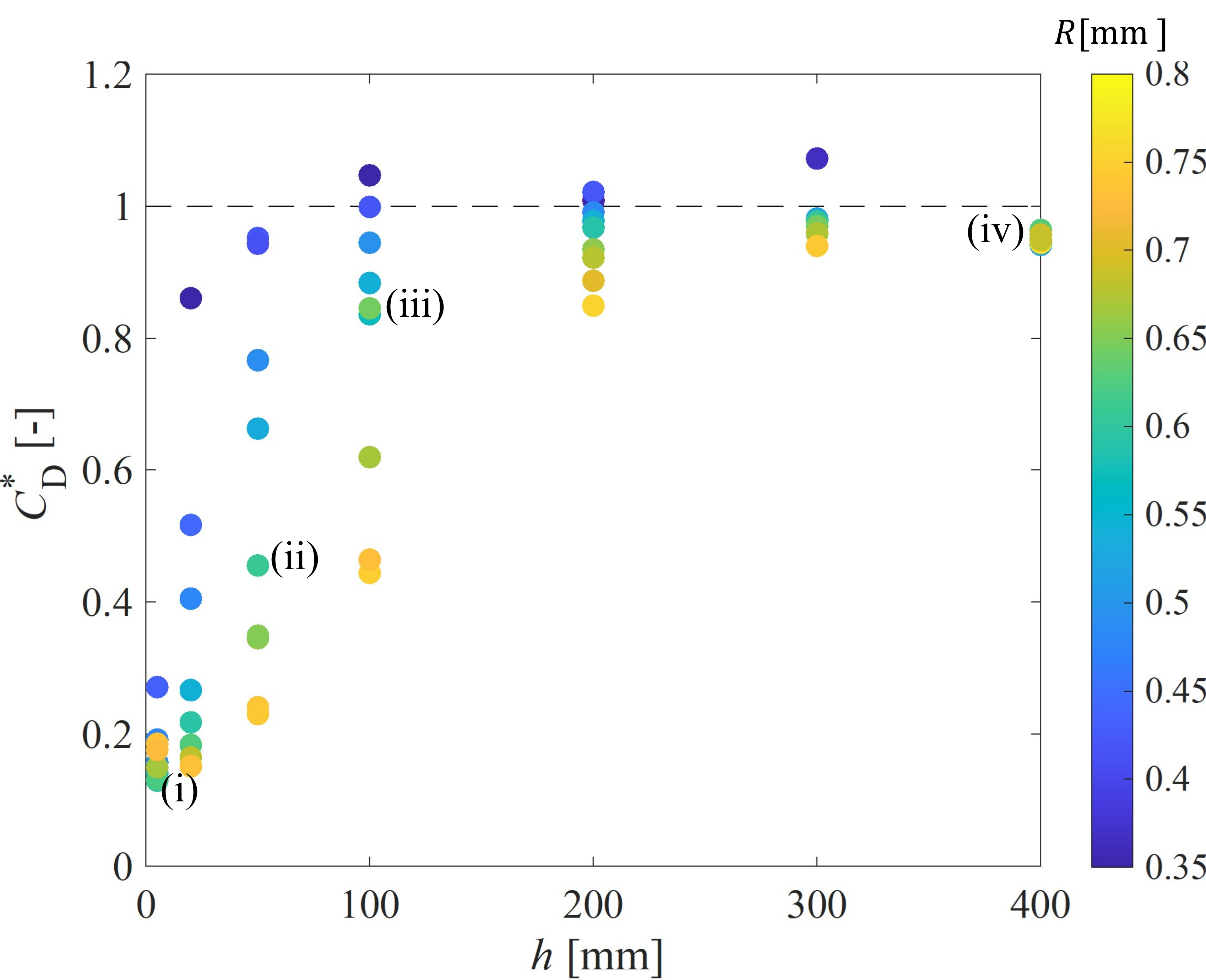}}
    \caption{Drag coefficient of the bubble for various rising heights. The colors show the bubble sizes in each experiment. (i)–(iv) correspond to the figure \ref{fig:main}. }
\label{fig:cd}
\end{figure}

\subsection{Flow around a contaminated bubble}\label{bubble}
\subsubsection{Surfactant-induced stress variation around a contaminated bubble}\label{experiment1}

We observed an evolution in the drag coefficient of bubbles rising in a surfactant solution (ultrapure water + CNC 0.5 wt$\%$ + Triton X-100 0.4 ppm).
Figure \ref{fig:cd} shows the drag coefficient ratio for various heights ($h=$ 5, 20, 50, 100, 200, 300 and 400 mm), with bubble radius in the range $0.35 < R < 0.8$ mm.
The drag coefficient ratio $C_D^*$ is expressed as
\begin{equation}
C_D^* = \frac{C_D-C_{D_b}}{C_{D_s}-C_{D_b}}\;,
\label{eq:cap}
\end{equation}
where $C_{D_b}$ is the drag coefficient of the clean bubble \citep{Mei1994}, and $C_{D_s}$ is the drag coefficient of a solid sphere \citep{cheng2009}.
The drag coefficient was estimated from the balance of buoyancy ($\rho V\mathbf{g}$), drag force ($\pi\rho C_D R^2 \mathbf{U|U|}/2$), and added mass force (-$\rho V/2\;\partial U/\partial t$).
Note that the size of the bubble was controlled in the experimental setup, but the camera was at fixed position and thus all results were for different individual bubbles.
As various drag coefficient models are available for solid spheres \citep{Goossens2019}, we had to choose our model carefully.
In the low $Re$ region, $Re \sim 20$ for the smallest bubble in this study, a discrepancy of up to 6 $\%$ was observed between the models.
In this study, the model proposed by \citet{cheng2009} was used as the drag coefficient model closest to the previous numerical results \citep{magnaudet1995} for $Re = 20$.
In the early stage of bubble displacement ($h=5$ mm), the drag coefficient of the bubbles in the CNC suspension was close to that of clean bubbles ($C_D^*= 0$), but it deviated slightly. 
For bubbles that traveled a longer distance, the drag force increased around a height of $h\sim$ 300 mm, the drag coefficient for all bubbles was closer to that of a solid sphere ($C_D^* = 1$).
We argue that the bubbles became progressively more contaminated and the transition of the boundary conditions occurred owing to the Marangoni effect.
The drag coefficient of some contaminated bubbles was even higher than that of solids.
This overshoot has been also observed in numerical results \citep{Cuenot1997,Kentheswaran2023}. 

\begin{figure}
    \centerline{\includegraphics[width=1.\linewidth]{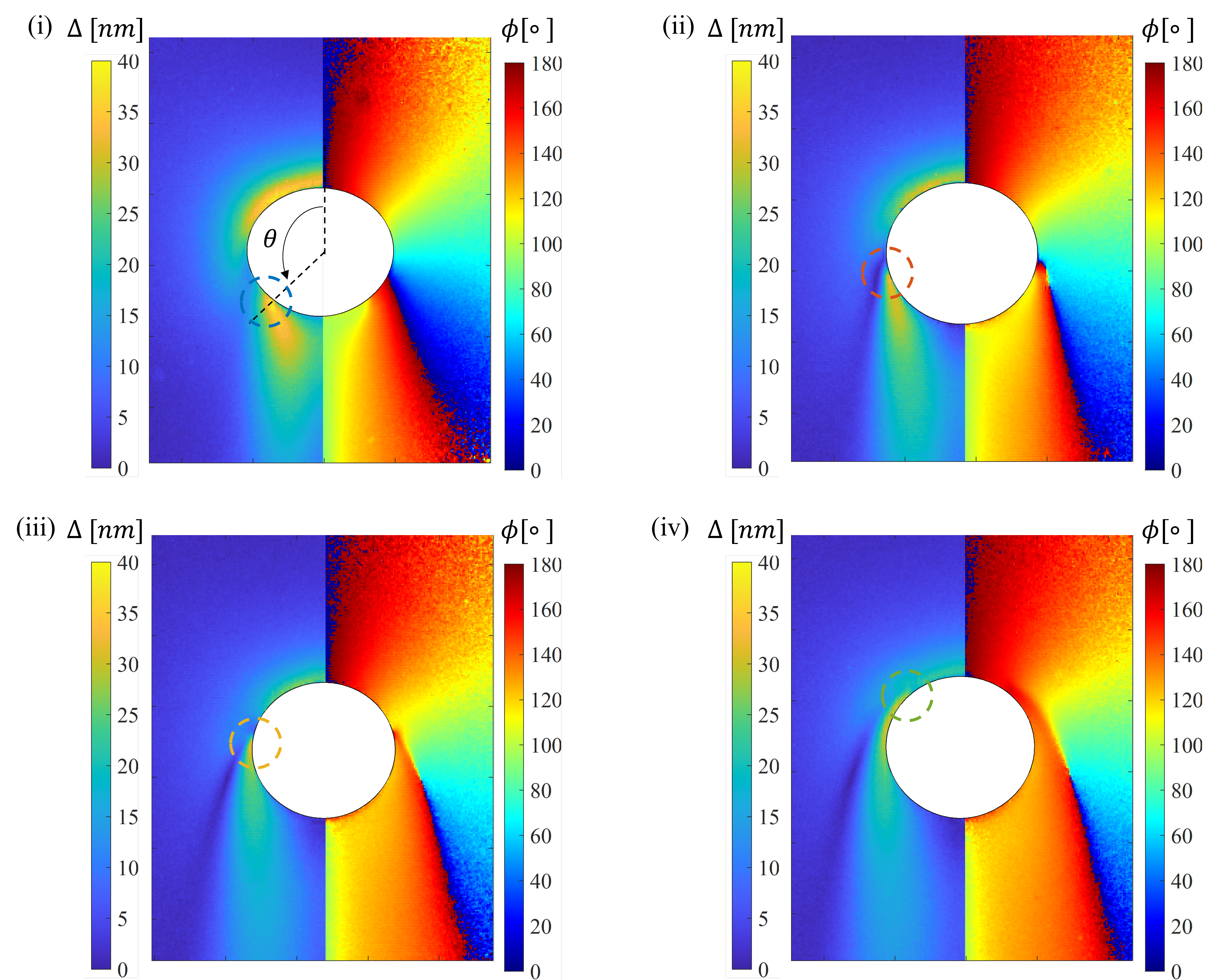}}
    \caption{Retardation (left) and azimuth (right) field around the bubble, whose radius was 0.59 $\pm$ 0.01 mm, at (i) $C_D$ = 0.39 ($h=5$ mm), (ii) $C_D$ = 0.69 ($h=50$ mm), (iii) $C_D$ = 1.18 ($h=100$ mm), (iv) $C_D$ = 1.22 ($h=400$ mm). The position at which retardation jump occurred is indicated by a dashed circle. The value of $C_D$ and $h$ are also indicated in figure \ref{fig:cd}}
\label{fig:main}
\end{figure}

\begin{figure}
    \centerline{\includegraphics[width=1\linewidth]{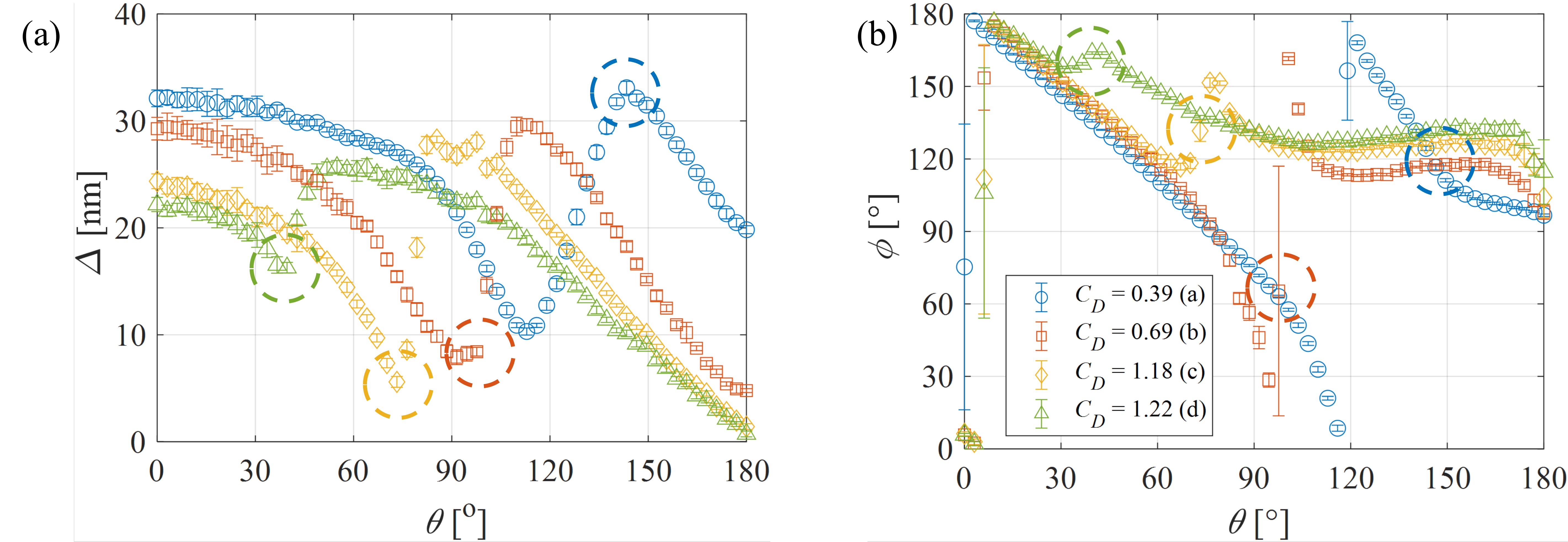}}
    \caption{(a) Phase retardation near the bubble surface ($r = 1.05R$). (b) Azimuth. The results for  $C_D$ = 0.39, 0.69, 1.18, and  1.22 correspond to those for  figure \ref{fig:main}(i)–(iv), respectively. The dashed circles indicate retardation jumps.}
\label{fig:main2}
\end{figure}

Let us examine the flow structure around the bubble at every height ($h=5,50,100$ and $400$ mm). 
Figure $\ref{fig:main}$(i)-(iv) show the polarization measurement results around the bubble with a diameter of 0.59 $\pm$ 0.01 mm at heights of $h$ = 5, 50, 100, and 400 mm, respectively, where the results are time averaged (5 ms, 10 frames).
Note that the averaging time was much shorter than the transient accumulation of the surfactant and much shorter than the bubble velocity change as discussed in \S\ref{sec:surfactants}.
The phase retardation and azimuth near the bubble ($r$ = 1.05$R$) are shown in figures \ref{fig:main2}(a) and (b).

A bubble at $h$ = 5 mm (figure \ref{fig:main}(i)) behaved closely to a clean bubble. 
First, for a nearly clean bubble (for more detail see figure \ref{fig:main2}(a)), the largest phase retardation was observed at the front stagnant point in the entire field. 
This indicated a completely different flow structure from the solid case, which had the largest phase retardation around the equator. 
The phase retardation was very small at the side of the bubble.
We observed that this was because the boundary condition was a free-slip condition.
Slightly further from the rear stagnant point ($\theta\sim$ 150$^\circ$), a large phase retardation was observed. 
This region was certainly a wake region, but the phase retardation was larger than in the solid sphere case.
Figure \ref{fig:main}(i) shows that little azimuth change occurred in the radial direction, which was not observed in the solid sphere in figure \ref{fig:solid}. 
In addition, the azimuth around the bubble exhibited near fore-aft symmetry, similar to a potential flow, indicating that the vorticity generated at the surface was small.
This indicated that the bubble surface largely maintained a free-slip condition, which resulted in a small deviation from the potential flow owing to insufficient vorticity generation. 
However, the azimuth deviated from the potential in the vicinity of the rear part indicated by the blue dashed circle.
Therefore, the reason for the deviation from the potential flow was considered to be a transition from free-slip to non-slip in the flow structure owing to contamination in the vicinity of the rear part.
This could be observed more clearly as the rise height increased with accumulating surfactants on the bubble surface.

The bubble at $h$ = 50 mm (figure \ref{fig:main}(ii)) was expected to be at intermediate contamination based on its drag coefficient. 
Because its rise velocity was smaller than that of the clean bubble, the magnitude of the phase retardation around the bubble at $h$ = 50 mm was smaller than that of the clean bubble because the phase retardation was proportional to the velocity in a similar flow, as can be observed from Eqs. (\ref{eq:Retadation1}) and (\ref{eq:Retadation2}). 
Nevertheless, in the front part of the bubble, both phase retardation and azimuth exhibited the same trend as that of a clean bubble. 
Note that in the region $\theta\sim105^\circ$, a jump occurred in the phase retardation and azimuth near the bubble surface, indicating that the boundary conditions have transitioned. 
The phase retardation and azimuth in the region below the jump were similar to those of the solid sphere, except the non-zero phase retardation at the rear stagnant point. 
At the rear of the bubble, we estimated that detachment occurred because $\phi$ increased with the increase in $\theta$.

The bubble at $h$ = 100 mm (figure \ref{fig:main}(iii)) had a drag coefficient similar to a solid sphere and could be assumed to be a ``fully contaminated'' bubble from the drag coefficient perspective.
The magnitude of the phase retardation was even smaller than for $h$ = 50 mm owing to the velocity reduction.
Now, in the region $\theta\sim70^\circ$, the phase retardation and azimuth changed drastically, i.e., a jump occurred near the bubble surface. 
The boundary condition was the non-slip condition in larger area than the bubble at $h$ = 50 mm because of the accumulated contamination, although the bubble was not a fully contaminated one from the flow structure perspective.
When the contamination advanced to this stage, the boundary condition for $0^\circ\le\theta\lesssim70^\circ$ was similar to that of the clean bubble in the front, and that for $70^\circ\lesssim\theta\le180^\circ$ was close to that of the solid sphere in the rear. 
Unlike the $h$ = 50 mm bubble, the phase retardation at the rear stagnation point was zero.
The fact that the drag coefficient agreed with that of the solid sphere although the entire flow field did not match that of the solid sphere can be attributed to the manner in which normal and shear stress contributions offset each other.
Specifically, although the local normal stress around $\theta \sim 70^\circ$ differed from that of a fully non-slip sphere, the front region ($0^\circ\le\theta\lesssim70^\circ$) retained a free-slip boundary condition, which balanced out this difference such that the overall normal force remained comparable to that of the solid sphere.
For the shear stress, its contribution to drag was weighted by $\sin\theta$.
Because $\sin\theta$ was small in the front part, the difference between free-slip and non-slip conditions had only a minor effect on the total shear contribution.
Consequently, although the local flow field deviated from the fully non-slip cases, the integrated normal and shear stresses produced nearly the same drag coefficient as that of a solid sphere, as indicated by previous numerical studies \citep[e.g.,][]{Cuenot1997}.

The bubble at $h$ = 400 mm (figure \ref{fig:main}(iv)) was the most contaminated bubble in this study. 
Compared with the bubble at $h$ = 100 mm, the boundary condition transition occurred more forward ($\theta\sim40^\circ$).
The evolution of the angle where the jump in the boundary condition occurred 30$^\circ$ ahead of the bubble of $h$ = 100 mm, although the rising trajectory was four times longer. 
This indicated that a very long time was required for the front part to become a non-slip condition.

The polarization measurement experiments described in this section showed that as the rising distance of the bubbles increased, the transition from the free-slip to non-slip condition (flow around a clean bubble to flow around a solid sphere) occurred in sequence from the rear of the bubbles. 
The point at which the boundary conditions transitioned appeared as a jump in the phase retardation and azimuth. 

At the point where the boundary condition shifts from free-slip to non-slip, a very strong vorticity (shear stress) must be generated at the surface to satisfy the continuity equation \citep{Cuenot1997}.

\begin{figure}
    \centerline{\includegraphics[width=1\linewidth]{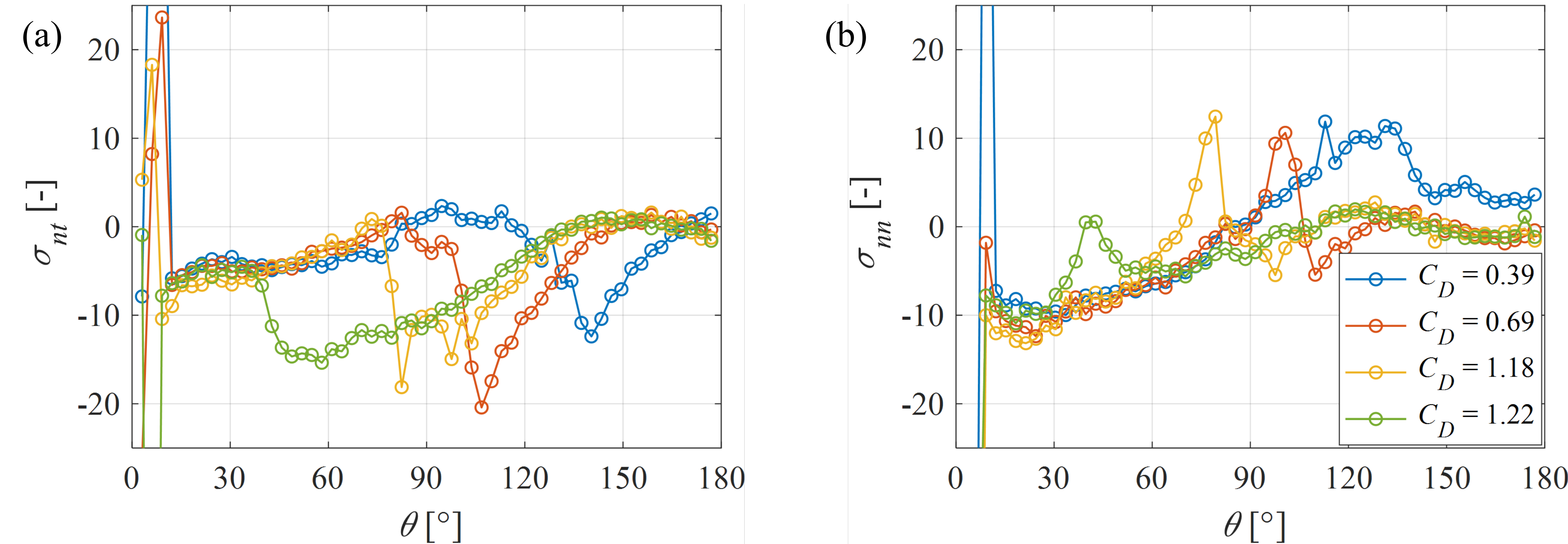}}
    \caption{(a) Normalized shear stress and (b) normalized normal stress near the bubble surface ($r = 1.05R$). Stress are normalized by $\mu U/2\rho R$. For conditions see figure \ref{fig:main}.}
\label{fig:stress}
\end{figure}

Here, we reconstructed the normalized shear stress $\sigma_{nt}$ and normal stress $\sigma_{nn}$ near the surface from the retardation and azimuth results (see appendix \ref{appA} for axisymmetric reconstruction method).
Note that the stress components are modified to the ($r$, $\theta$) coordinate along the surface from ($x$, $y$) coordinate system.
Figure \ref{fig:stress} (a) shows that a local large shear stress acted on the point where the stress jumped at each height. 
In the front, $\theta \sim 0^\circ$, artificial stresses oscillation occurred owing to reconstruction using the fluctuated azimuth.
In the front part except such artificial fluctuation region, a similar flow field was established irrespective of the bubble height. 
Figure \ref{fig:stress} (b) shows that the normal stress component also exhibited the spikes.
This was due to the non-slip behavior that caused the velocity to suddenly drop to zero.
In the free-slip condition, the shear stress was zero and the vorticity was $O(1)$ at the surface; therefore, such a large shear stress near the surface did not occur in a clean bubble.
Therefore, this revealed that the shear stress (or the vorticity proportional to it) was caused by the Marangoni stress \citep{Atasi2023}.
This local shear stress indicated the localized generation of a surface tension gradient (or surfactant concentration gradient), and we inferred that a small amount of surfactant was present in the forward part from the jump point.
Nevertheless, a small shear stress was still acting on the front part of the bubble, and the possibility of the effect of surfactant adsorption during advection could not be ruled out.


\begin{figure}
\centerline{\includegraphics[width=.7\linewidth]{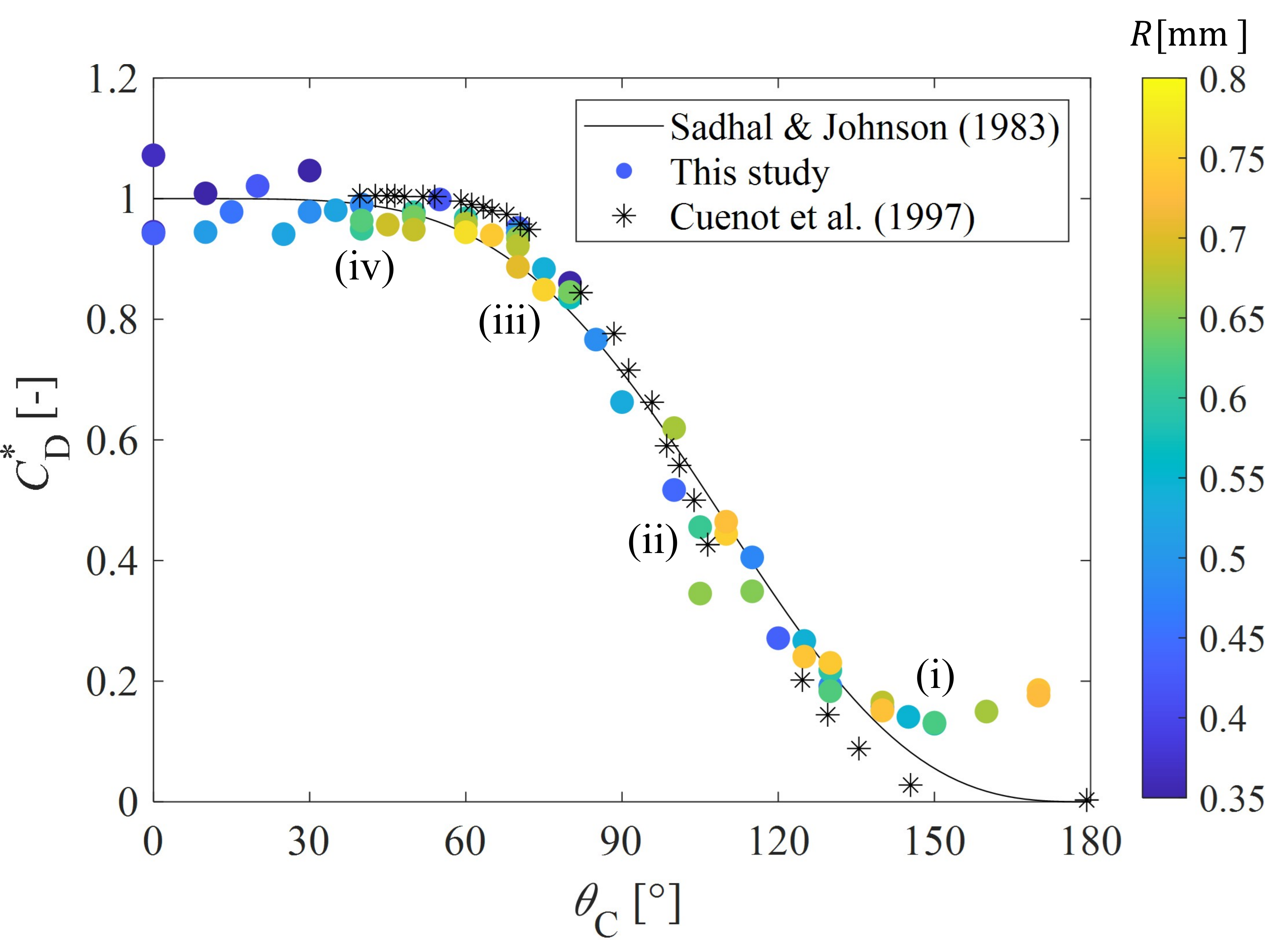}}
    \caption{Comparison of drag coefficient ratios of experimental results and stagnant cap model: ---$\!$---, stagnant cap model \citep{Sadhal1983a}; $\fullcirc$, experimental results (cap angle is obtained from phase retardation jump); $*$, numerical results \citep{Cuenot1997}. (i)-(iv) correspond to figure \ref{fig:main}.}
\label{fig:cap}
\end{figure}

\subsubsection{Measured cap angle comparison with the stagnant cap model and numerical simulations}\label{experiment2}

In this section, we compare the measured cap angle with both the stagnant cap model \citep{Sadhal1983a} and the numerical results from \citet{Cuenot1997}.
The stagnant cap model, which is often used for comparative verification of previous numerical analysis, has not been compared with experimental results.
Using the stagnant cap model , we can describe $C_D^*$ as a function of $\theta_C$ (see Eq. (53) of \citet{Sadhal1983a}):
\begin{equation}
C_D^*(\theta_C) = \frac{1}{2\pi}\left[
2(\pi-\theta_C) + \sin\theta_C+\sin2\theta_C-\frac{1}{3}\sin3\theta_C
\right].
\label{eq:sadhal}
\end{equation}
Figure \ref{fig:cap} shows the relationship between the normalized drag coefficient and cap angle. 
The experimental results for various bubble size exhibited reasonable agreement with the model overall.
This result strongly confirmed that the phase retardation and azimuthal jumps in the experiment corresponded to the shift of the boundary condition from free-slip to non-slip.
However, two discrepancies were observed.
The first was that the experimental drag at $140^\circ < \theta\rm_C$ was larger than the theory.
This was because the bubble had a spheroidal shape and the drag was larger than that of spherical shape \citep{blanco1995structure,sanada2008motion}. 
The second was that the slope of the experimental drag was slightly steeper than the theory near the equator ($\theta_C \sim 90^\circ$).
A similar trend was observed in the numerical analyses performed at a finite $Re$ of 100 \citep{Cuenot1997,Dani2022}, indicating that the drag transition is more sensitive than for the creeping flow.
To the best of the author's knowledge, this is the first experimental result that was consistent with previous numerical studies of the flow structure around bubbles in surfactant solutions with finite $Re$. 

Note that, although the drag coefficient of an apparently contaminated bubble approaches that of a solid sphere, the flow field differs, particularly for $\theta_C < 60^\circ$. 
Drag contributions primarily result from shear stress in the range $45^\circ < \theta_C < 135^\circ$ and normal stress elsewhere. 
As shown in figure~\ref{fig:stress}, viscous normal stress may counterbalance shear stress and pressure gradient, even when the bubble front remains clean. 
While this flow field difference likely has a slight impact on the drag coefficient, it can influence other phenomena, such as bubble path instability \citep{Tagawa2014,pesci2018computational,farsoiya2024coupled}, lift reversal \citep{Fukuta2008,Hayashi2018}, and stability of multi-bubbles \citep{Harper2008,Atasi2023}.


{\section{Conclusion}\label{conslusion}}
We experimentally determined the stress field of the flow around a bubble rising in a dilute surfactant solution using a newly developed unsteady polarization measurement method.
We first validated our method by measuring the flow around a solid sphere falling at terminal velocity in a quiescent solution without surfactant effect.
The measurements and numerical results were in quantitative agreement, confirming that the photoelastic method was successful in measuring the stress around the sphere.
The stress field around the solid sphere revealed distinct regions, including the outer potential flow region, boundary layer, and wake. 
Interestingly, the mushroom-shaped phase retardation observed at the front of the sphere was attributed to the interplay between potential flow characteristics and boundary layer dynamics.
These features, combined with azimuth profiles reflecting stress orientations, provided a detailed understanding of the stress distribution around the sphere and the mechanisms underlying flow structures at intermediate Reynolds numbers.

We then measured the flow around a bubble rising in a quiescent surfactant solution with a high Peclet number and intermediate Reynolds number. 
The results showed that the flow structure near the front of the bubble resembled that of a clean bubble, whereas the rear flow was similar to that of a solid sphere. 
However, the stress field in front of the bubble differed even when the drag coefficient was nearly the same, particularly when the cap angle was less than 60$^\circ$. 
Additionally, when the drag coefficient deviated from that of a solid sphere, slight differences in the rear stress field were observed. 
Sudden jumps in phase retardation and azimuth, indicated by the cap angle, were detected between the front and rear of the bubble. 
These jumps evolved with increasing distance traveled, reflecting the time-dependent adsorption of surfactants. 
By reconstructing the stress field under axisymmetric assumptions, we experimentally demonstrated the presence of localized stresses acting in the phase retardation jump region. 
The measured cap angle as a function of the normalized drag coefficient had reasonable agreement with the predictions of the stagnant cap model proposed by \citet{Sadhal1983a} and quantitatively aligned with numerical results at intermediate Reynolds numbers by \citet{Cuenot1997}, marking the first experimental confirmation of previous numerical studies on flow structures in the vicinity of a bubble in a dilute surfactant solution at finite Reynolds numbers.


As a future prospect, the method employed in this study may create new opportunities for revealing stress field in a wide range of dispersion flows.
Although only a simple system was used in this study, various systems exist in bubble dynamics, such as multi-bubbles and bubble motion in non-Newtonian fluids.
For low Peclet number conditions and mixed surfactants, the boundary conditions become half-slip, which is the slip condition between non-zero shear stress and non-zero velocity.
In addition, as the amount of vorticity generation increases under slip conditions, unstable behavior is caused by vorticity evacuation and accumulation generated at the surface.
Moreover, lift reversal \citep{Fukuta2008,Hayashi2018} and bubble train stabilization \citep{liger2000velocity,Harper2008,Atasi2023} are caused by vorticity stretching/tilting, respectively; both are organized by the maximum vorticity \citep{magnaudet2007,yang2007,cano2016}.
The polymer stretching a negative wake behind a bubble \citep{Zenit2018}.
Therefore, the extension of the results of this study to viscous stress, boundary layer region and wake regions will contribute to the elucidation of the detailed physical mechanism of these phenomena, as well as in complex flows.
\\\\
{\bf Acknowledgement.} The authors would like to thank Prof. Roberto Zenit for carefully reading the manuscript and for giving constructive comments.\\\\
{\bf Funding.} This work was supported by JSPS KAKENHI Grant Numbers JP22K20403, JP23K13251, and JP24H00289.
We also acknowledge support from The Ihara Science Nakano Memorial Foundation.\\\\
{\bf Declaration of interests.} The authors report no conflict of interest.\\

\appendix
\section{Axisymmetric reconstruction}\label{appA}
\begin{figure}
\centerline{\includegraphics[width=1\linewidth]{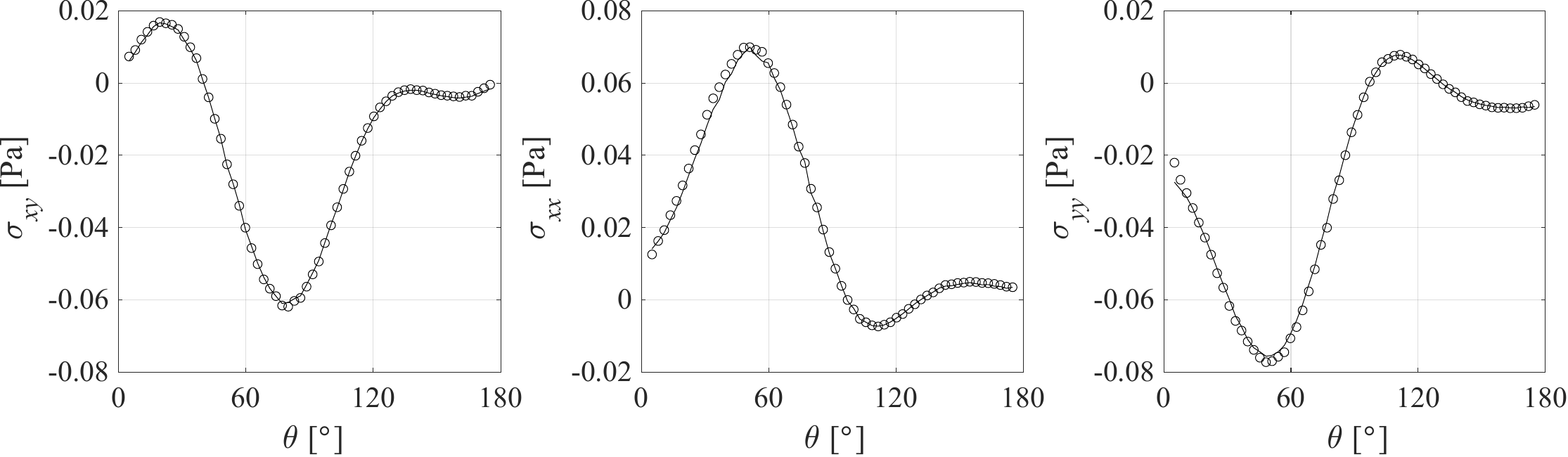}}
    \caption{Stress near the solid sphere surface ($r = 1.05R$);---$\!$---,Stress obtained from the velocity; $\bigcirc$, Stress obtained from the phase retardation and azimuth. For conditions see figure \ref{fig:solid}.}
\label{fig:reconst_test}
\end{figure}
In this section, we introduce the axisymmetric reconstruction procedure from the integrated phase retardation and azimuth to normal/shear stress.
Axisymmetric reconstruction has been conventionally used for transparent solids \citep{aben2010,Yokoyama2023}.
Inverse Abel transform and onion-peeling methods have been used for the reconstruction, and we adopted the onion-peeling method based on them.
For axisymmetric reconstruction, we must first transform the stresses in the Cartesian coordinate system to the cylindrical coordinate system $(r,y,\Theta)$, assuming a steady-state condition, as follows
\begin{equation}
\sigma_{xy} = \sigma_{ry}\cos\Theta\;,
\label{eq:cartesian_to_axis1}
\end{equation}
\begin{equation}
\sigma_{yy}-\sigma_{xx} = \sigma_{yy}-\sigma_{rr}\cos^2\Theta - \sigma_{\Theta\Theta}\sin^2\Theta\;.
\label{eq:cartesian_to_axis2}
\end{equation}
From Eq. (\ref{eq:Retadation2}), with the phase retardation and azimth obtained from the polarization measurement, the shear stress is independent of the other stress compornents, indicating that it can be estimated by a procedure similar to that previously used in the analysis of solids:
\begin{equation}
\Delta\sin2\phi=2C\int\sigma_{ry}\cos\Theta dz\;.
\label{eq:reconstruction1}
\end{equation}
In contrast, normal stress cannot be estimated in the same manner as for solids.
In the steady state of a solid, the derivation is based on the equilibrium equation $\nabla\cdot\boldsymbol\sigma = 0$, i.e., the shear and normal stresses are balanced, whereas no such limitations exists in a fluid owing to advection $(\mathbf{\nabla\cdot\boldsymbol\sigma} = \rho\mathbf{\nabla\cdot(uu)})$.
Therefore, we use the relation between the continuity equation and axisymmetric stresses assuming a Newtonian fluid.
We treat only the viscous stresses neglecting pressure because the pressure cancels out in Eq. (\ref{eq:cartesian_to_axis2}).
Assuming a Newtonian fluid, the continuity equation yields $\sigma_{rr}^{vis} + \sigma_{\Theta\Theta}^{vis} + \sigma_{yy}^{vis} = 2\mu(\partial u_r/\partial r+u_r/r + \partial u_y/\partial y) = 0$, where the subscript denotes the viscous stress compornent.
Additionally, the axisymmetric stress relation $\sigma_{rr}^{vis} = \partial(r\sigma_{\Theta\Theta}^{vis})/\partial r$, from Eqs. (\ref{eq:Retadation1}) and (\ref{eq:cartesian_to_axis1}), the relation between $\Delta$, $\phi$ and $\sigma_{\Theta\Theta}^{vis}$ becomes
\begin{equation}
\Delta\cos2\phi = -C\int
3\sigma_{\Theta\Theta}^{vis} + r(1+\cos^2\Theta)\frac{\partial\sigma_{\Theta\Theta}^{vis}}{\partial r}dz\;.
\label{eq:reconstruction2}
\end{equation}
When the $\Theta$ component is known, the other normal stress components are also uniquely determined.

We performed tests on numerical data (figure \ref{fig:solid}) to validate our reconstruction method.
Figure \ref{fig:reconst_test} shows the stress fields obtained from the velocity and phase retardation and azimuth fields using the onion-peeling method, respectively.
The stresses in the far from the symmetry axis were calculated as 0.
The figure shows that the stress fields obtained from the velocity and phase and azimuth were consistent over a wide range.
Therefore, we conclude that this reconstruction method is applicable.
Note that the values near the axis of symmetry and near the interface cannot be guaranteed owing to the orientation fluctuation as discussed in $\S\ref{sec:verification}$.

\bibliographystyle{jfm}
\bibliography{jfm-instructions}
\end{document}